\documentclass{article}

\usepackage{amssymb}
\usepackage{amsmath}
\usepackage{graphicx,placeins}
\numberwithin{equation}{section}

\newtheorem{theorem}{Theorem}[section]

\newtheorem{remark}[theorem]{Remark}

\newcommand{\ccz}{{\overline{z}}}

\newcommand{\ccV}{{\overline{V}}}

\newcommand{\cc}[1]{{\overline{#1}}}
\newcommand{\w}[1]{{\widetilde{#1}}}
\newcommand{\D}{\mathrm{d}}
\newcommand{\tr}{{\mathrm{tr}}}
\newcommand{\wW}{{\widetilde W}}
\newcommand{\uu}{{\underline{u}}}
\newcommand{\ux}{{\underline{x}}}
\newcommand{\atanh}{{\, \mathrm{atanh}}}
\newcommand{\pp}[2]{ \frac{\partial #1}{\partial #2} }
\begin{document}
\title{ Homogeneous Euler equation: blow-ups, gradient catastrophes and singularity of mappings}
\author{
B.G.Konopelchenko $^1$ and G.Ortenzi $^{2}$ 
\footnote{Corresponding author. E-mail: giovanni.ortenzi@unimib.it,  Phone: +39(0)264485725 }\\
$^1$ {\footnotesize Dipartimento di Matematica e Fisica ``Ennio De Giorgi'', Universit\`{a} del Salento, 73100 Lecce, Italy} \\
 $^2$ {\footnotesize  Dipartimento di Matematica e Applicazioni, 
Universit\`{a} di Milano-Bicocca, via Cozzi 55, 20126 Milano, Italy}\\
$^2${\footnotesize  INFN, Sezione di Milano-Bicocca, Piazza della Scienza 3, 20126 Milano, Italy}
} 
\maketitle
\abstract{
The paper is devoted to the analysis of the blow-ups of derivatives, gradient catastrophes and dynamics of mappings of 
$\mathbb{R}^n \to \mathbb{R}^n$ associated with the $n$-dimensional homogeneous Euler equation. Several characteristic features 
of the multi-dimensional case ($n>1$) are described. Existence or nonexistence of blow-ups in different dimensions, boundness of certain linear combinations of 
blow-up derivatives and the first occurrence of the gradient catastrophe are among of them. It is shown that the potential solutions of the
Euler equations exhibit blow-up derivatives in any dimenson $n$. Several concrete examples in two- and three-dimensional 
cases are analysed. Properties of  $\mathbb{R}^n_\uu \to \mathbb{R}^n_\ux$ mappings defined by the hodograph equations are studied,
including appearance and disappearance of their singularities. 
}
\section{Introduction}
The homogeneous Euler equation
\begin{equation}
\pp{u_i}{t}+\sum_{k=1}^n u_k \pp{u_i}{x_k}=0\, , \qquad i=1,\dots,n
\label{nBH}
\end{equation}
is an important and remarkable representative of the class of multidimensional quasi-linear partial differential equations. 
{It is the basic equation of the hydrodynamics and theory of continuous media , namely the Navier-Stokes equation, 
in the situation when one can neglect effects of pressure, dissipation, viscosity, dispersion etc (see e.g. \cite{L-VI,Lamb,Whi}). It can be viewed also as the inviscid multidimensional Burgers equation  (see e.g. \cite{SZ89,BK07})}.
In spite of such simplification, the equation (\ref{nBH}) arises in various branches of physics from hydrodynamics
to cosmology (see e.g. \cite{L-VI,Lamb,Whi,SZ89,Zel70}). \par

In addition, it has the remarkable property to be solvable 
 by multidimensional version of the 
classical method of hodograph equations \cite{SZ89,Zel70,Fai93,FL95}. Namely, any solution of equation (\ref{nBH}) is obtainable as a solution of the hodograph 
equations \cite{Fai93,FL95}
\begin{equation}
x_i -u_i t -f_i(\uu)=0\, , \qquad i=1,\dots,n
\label{nhodo}
\end{equation}
where $f_i(u_1,\dots,u_n)$ are arbitrary functions associated with the initial data $u_i(\ux,0)$ for equation (\ref{nBH}).\par

In virtue of all that the homogeneous Euler equation is an excellent touchstone for the study of various properties of multidimensional quasi-linear 
equations. \par

In the present paper we will study singularities associated with the homogeneous Euler equation, namely,
 blow-ups  of the derivatives $\pp{u_l}{x_k}$,  gradient catastrophes (blow-ups at $t>0$)  and the dynamics of singularities  
of the mappings $\mathbb{R}^n_\uu \to \mathbb{R}^n_\ux$ defined by the hodograph equations (\ref{nhodo}). These problems have been already 
partially addressed in \cite{SZ89,Zel70,Kuz03,KSZ94} using different techniques. Here the hodograph equations (\ref{nhodo}) will be our principal tool. \par

It is shown that all above singularities occur on the hypersurface in $R^{n+1}$ defined by the equation
\begin{equation}
t^n +\sum_{k=0}^{n-1}a_k(\uu)t^k=0
\label{singhyp}
\end{equation}
where coefficients $a_k(\uu)$ depend on the choice of the functions $f_k(\uu)$, $k=1,\dots,n$.    Blow-up (singularity) hypersurface (\ref{singhyp}) has 
$m$ branches where $m$ is the number of real roots of the equation (\ref{singhyp}). Classical property of the roots of polynomial equations with real 
coefficients imply that for odd $n=1,3,5,\dots$ there is always at least one branch of the hypersurface (\ref{singhyp}) while for even $n=2,4,\dots$
the minimal number of possible  real branches is zero. This means that for odd dimensions any solution of the equation (\ref{singhyp}) exhibits blow-up 
of the derivatives. Instead, in the even dimensional case, there are solutions of the Euler equations (corresponding to certain functions $f_k(\uu)$) 
free of blow-ups for real $t$. On the other hand, there are subclasses of solutions  for which blow-up always happens.  
It is shown that for the potential solutions ($u_i =\pp{\phi}{x_i}$, $i=1,\dots,n$) of the homogeneous Euler equation in any dimension $n$, the blow-up
hypersurface (\ref{singhyp}) has always $n$ real branches. Consequently, any potential solution of equation (\ref{nBH}) at any dimension $n$ exhibits
the blow-up of derivatives. \par

These multidimensional  features of blow-ups are illustrated by explicit examples in the two-dimensional case. 
It is shown that for the potential flow  which correspond to the functions 
$f_1=\pp{\wW}{x_1}$ and $f_2=\pp{\wW}{x_2}$ there are always two branches of 
singularity hypersurfaces. Hence, any solution from this subclass exhibits blow-up (for positive or negative $t$). \par

Opposite case can be easily analyzed by rewriting the two-dimensional Euler equations (\ref{nBH}) in the complex variables $z=x_1+ix_2$,
$V=u_1+iu_2$, namely, in the form $V_t+VV_z+\ccV V_\ccz=0$.
Under the reduction $V_\ccz=\mu(z,\ccz,t) V_z$, where $\mu$ is a certain function, it assumes the form 
\begin{equation}
V_t+(V+\mu \ccV) V_z=0\, ,
\label{2BHc-intro}
\end{equation}
plus the equation for $\mu$. It is shown that in the case $\mu \equiv 0$ ($V_\ccz=0$), i.e. for the complex Burgers-Hopf equation $V_t+VV_z=0$,
considered in \cite{KSZ94,KZ14,ZK18}, the equation (\ref{singhyp}) has no real roots. So, all such analytic solutions are blow-up free. For $\mu \neq 0$
 blow-up happens for $|\mu|=1$  that exactly coincides with the singularity of the quasi-conformal mapping given by the function $V(z,\ccz)$. \par
 
 In the paper we consider concrete examples of solutions of the two- and three-dimensional Euler equation (\ref{nBH}). Blows-ups and gradient catastrophes 
 for them are analysed with emphasis on differences with one-dimensional case. Among the characteristic properties of blow-up and gradient catastrophe (GC)
 for $n$-dimensional Euler equation we note two of them. First, one of the consequences of the equation (\ref{singhyp}) is that GC
 first happens at the point $(u_1,\dots,u_n)$  on the blow-up hypersurface (\ref{singhyp}) at the time $t_c$ and then expands on a whole blow-up
 hypersurface. Second property is the consequence of the degeneracy of a certain matrix $M$. Namely, even if all the derivatives $\pp{u_i}{x_k}$ 
 blow-up at the hypersurface (\ref{singhyp}), any their linear combinations in certain $(n-r)$-dimensional subspaces ($r=$rank$(M)$) remain finite.
 This property is manifestly a multi-dimensional feature of the Euler equation.\par
 
 {
 It is noted that the system (\ref{nhodo}) has an equivalent form, namely, 
\begin{equation} 
    x_i = {x_0}_i + {u_0}_i (\ux_0) t\, , \qquad  i=1,...n    \, ,
    \label{freepart-way}     
\end{equation}
where $x_i$ and ${x_0}_i$ are Eulerian and Lagrangian coordinates, respectively, and the functions ${u_0}_i$  (i=1,...,n) represent themselves the initial distribution of the components of the velocity \cite{Zel70} (see also \cite{Whi,SZ89,BK07,Kuz03}). In Zeldovich's theory \cite{Zel70} of the large scale structure of the universe (see also \cite{SZ89}) the system of equations (\ref{freepart-way}) describes the motion of the cold, collisionless medium  (dust).  It has been used also in the other models in physics (see e.g. \cite{BK07,Kuz03}). \par

   The results obtained in this  paper demonstrate that the hodograph equations in the form (1.2) are the simple and effective tool which allow us to perform most of the calculations explicitly up to the values of blow-ups times.\par
 }
 
 In the paper we also consider the dynamics of the mapping $\mathbb{R}^n_\uu \to \mathbb{R}^n_\ux$ defined by the hodograph equations (\ref{nhodo}). 
 It is shown that these mappings are singular on the hypersurfaces (\ref{singhyp}) with particular degenerations of these mappings around any point
 on the hypersurface (\ref{singhyp}). Appearance and disappearance of singularities and their alternation are analyzed for some concrete cases.
 In particular, it is shown the classical stable mappings  $\mathbb{R}^2 \to \mathbb{R}^2$ and $\mathbb{R}^3 \to \mathbb{R}^3$ \cite{Whi55,AGV}, i.e. folds, cusps 
 and swallow tails remain singular on certain hypersurfaces at any $t$.\par 
 
 The paper is organized as follows. General properties of blow-ups and gradient catastrophe for $n$-dimensional homogeneous Euler equations are studied
 in Section \ref{sec-blowup}. The mappings $\mathbb{R}^n \to \mathbb{R}^n$ associated with the Euler equation are considered in section \ref{map-sec}.
 In Section \ref{potential-sec} the blow-ups of the potential solution of the Euler equation are discussed.
 Section \ref{2Dcase-sec} contains the general discussion of the two-dimensional case. Concrete examples of the solutions of the homogeneous 
 Euler equation, their gradient catastrophes and all that are considered in section \ref{2Dexa-sec} for 2D cases, in section \ref{mapexe-sec} for dynamics of 
 mappings case and finally in section \ref{3Dexe-sec} for 3D cases.  
 In the conclusion (section \ref{conclusion-sec})  some possible directions of future investigations are indicated.
\section{Blow-ups and gradient catastrophes for n-dimensional Euler equation} \label{sec-blowup}
Let us start with the classical textbook case (see e.g. \cite{Whi}) of the one-dimensional Euler equation (Burgers-Hopf-Riemann equation).
Hodograph equation is given by ($u\equiv u_1)$
\begin{equation}
x=ut+f(u)\, ,
\label{1hodo}
\end{equation}
where $f(u)$ is the function locally inverse to the initial data $u_0(x)=u(t=0,\ux)$. Differentiating (\ref{1hodo}) w.r.t. $x$ and $t$,
and assuming that $t+\pp{f}{u}\neq 0$, one gets
\begin{equation}
u_x=\frac{1}{t+\pp{f}{u}}\, , \qquad u_t=-\frac{u}{t+\pp{f}{u}}\, .
\label{uder1D}
\end{equation}
Consequently, any solution of the equation (\ref{1hodo}) is a solution of the Burgers-Hopf (BH) equation.\par

For a given function $f(u)$, i.e. given initial data $u_0(x)$, the derivatives blow up at $t=t_0$ defined by the equation
\begin{equation}
t+\pp{f}{u}=0\, .
\label{1catcur}
\end{equation}
If  $t_0>0$, then the blow-up of $u_x$ and $u_t$ is usually called  ``gradient catastrophe'' (see e.g. \cite{Whi}).
Otherwise will refer to such a situation as  ``blow-up'' of derivatives. \par

We note that in the one-dimensional case any solutions of the Euler equation (\ref{nBH}) exhibits a blow-up of derivatives $u_x$, $u_t$ (at negative
or positive $t_0$) while the gradient catastrophe ($t_0>0$) occurs only for certain initial data. We will see 
that at $n \geq 2$ the situation is quite different. \par

In the $n$-dimensional case, the hodograph equation (\ref{nhodo}) imply that (see also \cite{Fai93,FL95}) 
\begin{equation}
\begin{split}
& \sum_{l=1}^n M_{il} \pp{u_l}{x_k}= \delta_{ik}\, , \qquad i,k=1,\dots,n,\\
& \sum_{l=1}^n M_{il} \pp{u_l}{t}= -u_i\, , \qquad i=1,\dots,n, 
\end{split}
\label{udersys}
\end{equation}
where
\begin{equation}
M_{il}=\pp{f_i}{u_l}+t \delta_{il}\, , \qquad
i,l=1,\dots,n \, .
\label{Mdef}
\end{equation}
Assuming that $\det M \neq 0$, one gets
\begin{equation}
\begin{split}
\pp{u_i}{x_k}&=(M^{-1})_{ik}\, , \qquad  i,k=1,\dots,n,\\
\pp{u_i}{t}&=-\sum_k (M^{-1})_{ik}u_k\, , \qquad  i=1,\dots,n,
\end{split}
\label{uder}
\end{equation}
and, hence, any solution $u_i$, $i=1,\dots,n$ of the hodograph equation (\ref{nhodo}) with $\det M \neq 0$ obeys the Euler equation (\ref{nBH}).\par
In the hodograph equation (\ref{nhodo}), the functions $f_i(\uu)$ are local inverse to the initial values $u_i(\ux,t=0)={u_0}_i(\ux)$, and the solutions of the 
homogeneous Euler equation (\ref{nBH}) are given implicitly by the formula \cite{Fai93,FL95}
\begin{equation}
u_i(\ux,t)={u_0}_i(\ux-\uu t)\, , \qquad i=1,\dots, n\, .
\label{dirsolnBH}
\end{equation}
This formula implies that the domain $\mathcal{D}_\uu \subset \mathbb{R}^n$ of variations of the functions $u_i(\ux,t)$ coincides with the domain 
of variation of the initial data  ${u_0}_i(\ux)$.\par

{We note that generically the correspondence between the initial data ${u_0}_i (x)$ and the functions $f_i (u)$ is not, obviously, one-to-one. Similar to the one-dimensional case one may have several functions $f_i(u)$ for the given initial data ${u_0}_i (x)$, one for every open set of invertibility of the initial datum. So, for the given initial data ${u_0}_i (x)$  $(i=1,...,n )$ one may have several associated matrices M of the form (\ref{Mdef}). }

The matrix $M$ is the central ingredient in this construction. It is easy to show that it obeys the equation
\begin{equation}
\frac{\D M}{\D t} =E\, ,
\label{matMsys}
\end{equation}
where $\frac{\D }{\D t} =\pp{}{t}+ \sum_{k=1}^n u_k \pp{}{x_k} $ and $E$ is the identity matrix. For the matrix $M^{-1}$ one has the equation ($M^{-1}\equiv U$)
\begin{equation}
\frac{\D U}{\D t} +U^2=0\, ,
\label{matUsys}
\end{equation}
that is the equation used in \cite{Kuz03} . \par

The matrix $M$ is also the key object in the analysis of blow-ups and gradient catastrophes (GC) for the equation (\ref{nBH}). The formulas (\ref{udersys}) and
(\ref{uder}) imply that blow-up of derivatives occurs when 
\begin{equation}
\det M =0\, .
\label{catMcond}
\end{equation}
Due to (\ref{Mdef}) this condition is of the form
\begin{equation}
t^n+a_{n-1}(\uu)t^{n-1}+ \dots + a_0(\uu)=0\, ,
\label{surcatpoly}
\end{equation}
where the coefficients $a_k(\uu)$ are certain real-valued functions of $\uu=u_1, \dots, u_n$ and $a_0(\uu)=\det \pp{f_i}{u_k}$.
We will refer to the hypersurface in $\mathbb{R}^{n+1}_{(t,\uu)}$ defined by the equations (\ref{surcatpoly}) as the blow-up hypersurface. 
Its structure and properties depend on the solution.
 {For the given initial data ${u_0}_i$ one may have several equations of the form (\ref{surcatpoly})  with different coefficients $a_k (u)$, $k =0,..., n-1$. 
 It is important that all of them have the same order $n$.}
 In general the blow-up hypersurface has $m$ branches
\begin{equation}
t=\bigcup_{i  \in \mathcal{S}} t_i(\uu)\, , \qquad \mathrm{with} \quad t_i(\uu)= \phi_i(\uu)\, , \quad i  \in \mathcal{S}\, ,
\label{tbranches}
\end{equation}
corresponding to $m$ real roots  ${t}_i$ of the polynomial equations (\ref{surcatpoly}) where $\mathcal{S}$ is the subset of indices $i$ for which
${t}_i$ is real. \par

Due to the standard properties of the roots of polynomial with real coefficients the maximal number of branches (\ref{tbranches}) is equal to $n$.
Minimal number of branches (\ref{tbranches}) is equal to one fo odd $n$ and to zero for even $n$. This means that for $n=1,3,5,\dots$ each solution 
of the Euler equation exhibits a blow-up while for $n=2,4,6, \dots$ there are solutions free of blow-ups. GC appears on the  branches $f_i(\uu)$ for which
${t_0}_i>0$. Several examples of solutions at $n=2,3$ 
with different properties will be presented in subsequent sections. \par

General properties of the branches  (\ref{tbranches}), i.e. their coalescence and intersections, are loosely connected with  the well-known problem 
of the stratifications of the space of matrices \cite{AGV,Arn71}. In our case it is the family of $n \times n$ matrices $M$ 
(\ref{Mdef}) with the parameter $t$.
 The condition that $s$  branches (\ref{tbranches}) coalesce is equivalent to $s-1$ partial differential equations
 \begin{equation}
\phi_i=\phi_k\, , \qquad i,k \in \mathcal{S} \, ,
 \end{equation}
for the functions $f_1,\dots,f_n$. On the other hand two branches $\phi_\alpha(\uu),\phi_\beta(\uu)$ (\ref{tbranches}) for given $f_1,\dots,f_n$ generally intersect along 
$(n-1)$-dimensional hypersurfaces  $\phi_\alpha(\uu)=\phi_\beta(\uu)$. \par
 
In the particular case $\pp{f_i}{u_k}=0$, $i \neq k $, i.e. $f_i=f_i(u_i)$, $i=1, \dots,n$ the $n$-dimensional equation (\ref{nBH}) decomposes into $n$ one-dimensional 
BH equations for the pairs of variables $(x_m,u_m)$, $m=1, \dots,n$. In this case the blow-up hypersuface has $n$ branches
\begin{equation}
t+\pp{f_m(u_m)}{u_m}=0\, , \qquad m=1, \dots,n\, .
\label{decnBH}
\end{equation}
The $m$-th branch (\ref{decnBH}) is associated with the BH equation for the variables $x_m,u_n$ and represent itself a cylindrical hypersurface 
generated by the curve (\ref{decnBH}).\par 

In physical problems the first time of appearance of GC (minimal value of $t$) is usually of most interest (see e.g. \cite{Whi}). Let the branch for which $t$
assumes the minimal value among the other is given by 
\begin{equation}
t_c=\phi(\uu)\, .
\label{tc-surf}
\end{equation}
The minimal value of such $t_c$ is defined by the condition
\begin{equation}
\pp{t_c}{u_i}=\pp{\phi(\uu)}{u_i}=0\, , \qquad i=1,\dots,n\, ,
\label{tcgrad0}
\end{equation}
plus a condition on the second derivatives. \par

For generic initial data the function $\phi(\uu)$ is a generic one.  Consequently, $n$ equation (\ref{tcgrad0}) has generically a single solution $\uu$. 
Thus, generically, the GC for the homogeneous Euler equation (\ref{nBH}) first happens at the time 
\begin{equation}
{t_c}_{\mathrm{min}}= \phi(\uu_c)
\label{cattc}
\end{equation}
at the point $\uu_c$ on the hypersurface (\ref{tc-surf}). Then it expands on the whole hypersurface (\ref{tc-surf}). 
It is noted that for the first time such property of the GC for multi-dimensional PDEs has been observed in \cite{Kuz03,MS08,MS11}. \par

Let us turn back to blow-ups. Due to the  formulae (\ref{uder}) all derivatives of $u_i$ blow-up simultaneously at the blow-up hypersurface (\ref{tc-surf})
similar to the case $n=1$ (\ref{uder1D}). In the multi-dimensional case the blow-ups and GC exhibit additional and novel properties.
Indeed, the first of the relations (\ref{uder}) can be equivalently rewritten as
\begin{equation}
\pp{u_i}{x_k}=\frac{\widetilde{M}_{ik}}{\det M}\, , \qquad i,k=1,\dots,n\, ,
\end{equation}
where $\widetilde{M}$ is the adjugate matrix. In vicinity of the blow-up hypersurface (\ref{surcatpoly}) one has
\begin{equation}
\det{M(t=t_0+\epsilon)}=\epsilon^n+A_{k-1}(\uu)\epsilon^{n-1}+\dots+A_1\epsilon\, ,
\end{equation}
since $\det M(t=t_0)=0$. On the other hand 
\begin{equation}
\widetilde{M}_{ik}(t=t_0+\epsilon)=\widetilde{M}_{ik}(t_0)+\epsilon \widetilde{M}_{ik}'(t_0)+\dots\, .
\label{M-near-tc}
\end{equation}
So, at $\epsilon \to 0$ one has
\begin{equation}
\pp{u_i}{x_k} =\frac{\widetilde{M}_{ik}(t_0)}{A_1} \frac{1}{\epsilon}+O(1)\, , \qquad i,k=1, \dots, n\, .
\label{udereps}
\end{equation}
Thus, at $\epsilon \to 0$, all $\pp{u_i}{x_k} \to \infty$. However, since the matrix $M$ is degenerate one, the matrix $\widetilde{M}$ is degenerate too.
Let the rank of the matrix $\widetilde{M}$ be equal to $r$. So, there are $n-r$ real vectors 
$\vec{\widetilde{R}}^{(\alpha)}=(\widetilde{R}^\alpha_1,\dots,\widetilde{R}^\alpha_n)$ and $n-r$ vectors
$\vec{\widetilde{L}}^{(\alpha)}=(\widetilde{L}^\alpha_1,\dots,\widetilde{L}^\alpha_n)$, $\alpha=1,\dots,n-r$ 
such that
\begin{equation}
\sum_{k=1}^n \widetilde{M}_{ik}(t_0)\widetilde{R}^{(\alpha)}_k=0\, ,\qquad \alpha=1,\dots,n-r\, , \quad i=1,\dots,n\, ,
\label{reigen}
\end{equation}
 and 
 \begin{equation}
\sum_{i=1}^n \widetilde{L}^{(\alpha)}_i \widetilde{M}_{ik}(t_0)=0\, ,\qquad \alpha=1,\dots,n-r\, , \quad k=1,\dots,n\, .
\label{leigen}
\end{equation}
These imply that
 \begin{equation}
\sum_{\alpha=1}^{n-r} \sum_{k=1}^n a_\alpha \pp{u_i}{x_k}  \widetilde{R}^{(\alpha)}_k \sim O(1)\, ,\qquad i=1,\dots,n\, , 
\label{rsupeigen}
\end{equation}
and
 \begin{equation}
\sum_{\beta=1}^{n-r} \sum_{k=1}^n  b_\beta \widetilde{L}^{(\beta)}_i \pp{u_i}{x_k} \sim O(1)\, ,\qquad k=1,\dots,n\, ,
\label{lsupeigen}
\end{equation}
where $a_\alpha$ and $b_\beta$ are arbitrary constants. 
Thus, at $n \geq 2$ derivatives $\pp{u_i}{x_k}$ all blow-up at the blow-up hypersurface. However, there are $(n-r)$-dimensional subspaces where 
all linear superpositions of derivatives (\ref{rsupeigen}), (\ref{lsupeigen}) are finite. \par

If $A_1=0$ then, instead of (\ref{udereps}), one has
\begin{equation}
\pp{u_i}{x_k} \sim \frac{1}{\epsilon^2} \frac{\widetilde{M}_{ik}(t_0)}{A_2} \qquad \mathrm{when} \quad\epsilon \to 0\, ,
\end{equation}
and so on. Thus, the first order blow-up sector for the $n$-dimensional Euler equation has a specific fine sctructure in contrast to the $n=1$ case of 
Burgers-Hopf equation. Higher order blow-up sectors and higher order GCs analogous for those in one-dimensional case \cite{KK02}  will be considered 
elsewhere. \par

The matrix $M$ serves also to define the type of behavior of the derivatives $\pp{u_i}{x_k}$ near the blow-up hypersurface. Indeed, considering the infinitesimal variations 
of $x_i$, $u_i$ for the fixed $t=t_0$ in the formula (\ref{nhodo}) one gets
\begin{equation}
\delta x_i = \sum_{k=1}^n M_{ik}(t_0) \delta u_k + \frac{1}{2} \sum_{k,l=1}^n\frac{\partial^2 f_i}{\partial u_k \partial u_l} \Big|_{t_0} \delta u_k \delta u_l +\dots\, , \qquad
i=1,\dots,n\, . 
\label{xdevu}
\end{equation}
Due to the degeneracy of the matrix $M(t_0)$ of rank $r$, there are $n-r$ vectors $\vec{W}^{(\alpha)}$ and $\vec{\widetilde{W}}^{(\alpha)}$ such that
\begin{equation}
\sum_{k=1}^n M_{ik}(t_0) {R}^{(\alpha)}_k=0\, , \qquad i=1,\dots,n\, , \quad \alpha=1,\dots,n-r\, ,
\label{reM}
\end{equation}
and 
\begin{equation}
\sum_{i=1}^n {L}^{(\alpha)}_i M_{ik}(t_0) =0\, , \qquad k=1,\dots,n\, , \quad \alpha=1,\dots,n-r\, .
\label{leM}
\end{equation}
Using (\ref{xdevu}) and (\ref{leM}), one obtains
\begin{equation}
\sum_{i=1}^n {L}^{(\beta)}_i \delta x_i = \sum_{k,l=1}^n \w{\phi}^\beta_{kl} \delta u_k \delta u_l+ \dots\, , \qquad \beta=1,\dots,n-r\, , 
\end{equation}
where
\begin{equation}
\w{\phi}^\beta_{kl}=\frac{1}{2} \sum_{i=1}^n \frac{\partial^2 f_i }{\partial u_k \partial u_l}\Big{\vert}_{t_0} {L}^{(\beta)}_i\, .
\end{equation}
In the case of nondegeneracy of the matrices $\w{\phi}^\beta_{kl}$, the formula (\ref{xdevu}) defines  generically the variation $\delta u_k$ as the square roots 
of the variation $\delta u_k$ as the square toots of the variations
\begin{equation}
\delta \xi^\beta=\sum_{i=1}^n {L}^{(\beta)}_i \delta x_i\, , \qquad \beta=1,\dots,n-r\, .
\end{equation}
The formula (\ref{xdevu})  has another consequence. If one consider the variation of $u_k$ of the form
\begin{equation}
\delta u_k=\sum_{i=1}^{n-r} {L}^{(\alpha)}_k \delta a_\alpha\, ,
\label{uvar}
\end{equation}
where $\delta a_\alpha$ are arbitrary infinitesimals, then the formula (\ref{xdevu}) implies that 
\begin{equation}
\delta x_i = \sum_{\alpha,\beta=1}^{n-r} {\phi}^{\alpha \beta}_{i} \delta a_\alpha \delta a_\beta \, , \qquad i=1,\dots,r\, , 
\label{xvar}
\end{equation}
where
\begin{equation}
{\phi}^{\alpha \beta}_{i}=\frac{1}{2} \sum_{k,l=1}^n \frac{\partial^2 f_i }{\partial u_k \partial u_l}\Big{\vert}_{t_0}  {L}^{(\alpha)}_k {L}^{(\beta)}_l\, .
\end{equation}
The formula (\ref{uvar}) together with (\ref{xvar}) also defines the behavior of $\delta u_i$ as a function of $\delta x_i$.
\section{Euler equation and mappings $\mathbb{R}^n \to \mathbb{R}^n$} \label{map-sec}
There are at least three ways to treat the mappings associated with the homogeneous Euler equations (\ref{nBH}).  \par

The first one, most standard, is to consider the family of mappings $\mathbb{R}^n_{(\ux)} \to \mathbb{R}^n_{(\uu)}$ given by a solution 
\begin{equation}
u_i=u_i(x_1,\dots,x_n)\, , \qquad i=1,\dots,n\, ,
\label{mapBH}
\end{equation}
of equation (\ref{nBH}). The Jacobian  $J_{(\ux,\uu)}=|\partial u_i/ \partial x_k| $ of these mappings is (see (\ref{udersys}), (\ref{matUsys}))
\begin{equation}
J= \det M^{-1}=\det U\, .
\label{JacM}
\end{equation}
One can show (see \cite{Kuz03}) that
\begin{equation}
U=U_0(1+U_0)^{-1}\, ,
\end{equation}
where ${U_0}_{ik}= \partial u_i/ \partial x_k |_{t=0}$. Hence ($J_0=\det U_0$)
\begin{equation}
J_{(\ux,\uu)}=\frac{J_0}{\det(1+U_0t)}\, .
\label{Jsol}
\end{equation}
The mapping (\ref{mapBH}) is singular if $J_{(\ux,\uu)}=0$ that defines the hypersurface
\begin{equation}
J_{(\ux,\uu)}(\ux)=0\, ,
\end{equation}
in the space $\mathbb{R}_{(\ux)}$ with local coordinates $\ux=x_1,\dots,x_n$. The formula (\ref{Jsol}) implies that the mapping (\ref{nBH}) singular at $t=0$
($J_0$) remains singular at all $t$, except (possibly) those values of $t_0$ for which $\det(1+U_0t)=0$. Such  $t_0$ corresponds to the blow-ups 
of the derivatives $\partial u_i/ \partial x_k$ for non-singular  $J_0 \neq 0$ mappings (\ref{nBH}) \cite{Kuz03}. \par

Second way is to consider the mappings $\mathbb{R}^n_{(\ux_0)} \to \mathbb{R}^n_{(\ux)}$ described by the version of the hodograph equation discussed in 
\cite{SZ89,Zel70}, namely, by equations
\begin{equation}
x_i ={x_i}_0 +v_i(\ux_0)t\, , \qquad i=1,\dots,n\, ,
\label{Zelhodo}
\end{equation}
where $v_i(\ux_0)$ is the  Lagrangian velocity of a particle starting from $\ux_0$. The Jacobian $J_{(\ux_0,\ux)}$ of such mapping is
\begin{equation}
J_{(\ux_0,\ux)}=\det(\delta_{ik}+ {U_0}_{ik} t) \, ,
\label{ZelJac}
\end{equation}
The mapping (\ref{Zelhodo}) is singular when 
\begin{equation}
\det(\delta_{ik}+{U_0}_{ik} t) =0\, .
\label{Zelcat}
\end{equation}

In this paper we will consider families of mappings $\mathbb{R}^n_u \to \mathbb{R}^n_x$ defined by the hodograph equations(\ref{nhodo}), namely, by
\begin{equation}
x_i=u_i t +f_i(\uu)\, , \qquad i=1,\dots,n\, .
\label{nhodomap}
\end{equation}
For given initial data ${u_0}_i(\ux)$ the formula (\ref{nhodomap}) defines the mapping of the domain $\mathcal{D}_\uu$ to the corresponding 
domain $\mathcal{D}_\ux$.

The Jacobian of these mappings is
\begin{equation}
J_{(\uu,\ux)}=\det M\, ,
\end{equation}
where the matrix $M$ is given by (\ref{Mdef}).
The family of mappings (\ref{nhodomap})  can be viewed as the deformations of the initial mappings
\begin{equation}
x_i=f_i(\uu)\, , \qquad i=1,\dots,n\, .
\label{ininhodomap}
\end{equation}
with the simple deforming part $u_i t$. \par

The mappings (\ref{nhodomap}) are singular ($J_{(\uu,\ux)}=0$) on the hypersurfaces in the space with the coordinates $(\uu,t)$ given by 
equation (\ref{surcatpoly}). Singularities of the mappings (\ref{nhodomap}) are obviously (as in many other cases) in one-to-one correspondence 
with blow-ups for the derivatives $\pp{u_i}{x_k}$. \par

Properties of blow-ups discussed in the previous section have their counterparts for the mappings (\ref{nhodomap}). Indeed, considering the 
infinitesimal variation $\delta u_i$ in vicinity of singular hypersurface, one has
\begin{equation}
\delta x_i =\sum_{k=1}^n M_{ik}(0) \delta u_k \, , \qquad i=1,\dots,n\, .
 \end{equation}
Hence, for the variation of $u_k$ given by 
\begin{equation}
\delta u_k= \sum_{\alpha_1}^{n-r} \delta a_\alpha {R}^{(\alpha)}_k\, ,
\end{equation}
where the vectors $R^{(\alpha)}_k$ are defined in (\ref{reM}), $\delta a_\alpha$, $\alpha=1,\dots,n-r$ are arbitrary infinitesimals, one has
\begin{equation}
\delta x_i=O(1)\, , \qquad i=1,\dots,n\, .
\end{equation}
Thus, around any point on the singularity hypersurface there is an $(n-r)$-dimensional infinitesimal domain whose image under the mapping
(\ref{nhodomap}) collapses to zero. \par

Analogously, due to the existence of $n-r$ vectors ${L}^{(\beta)}_i$, $\beta=1,\dots,n-r$, defined in (\ref{reM}),
one concludes that
\begin{equation}
\sum_k {L}^{(\beta)}_i \delta x_i  =o(|\delta u|^2)\,  .
\end{equation}
So, any $n$-dimensional infinitesimal domain around a point on the singularity hypersurface is transformed by the mapping $\ref{nhodomap}$
into the $r$-dimensional infinitesimal domain in the space $\mathbb{R}_{\uu}$.\par

Family of mappings (\ref{nhodomap}) can be viewed as that describing dynamics of the initial mapping (\ref{ininhodomap}) in time.
Study of the appearance or disappearance of singularities of mapping or their alternation is, definitely, of interest. 
Such properties of the mapping (\ref{ininhodomap}), obviously, are connected with the properties of stable and unstable mappings 
$\mathbb{R}^n \to \mathbb{R}^n$ studied by Whitney and others \cite{Whi55,AGV}. \par
In particular, mappings with stable singularities at $t=0$ 
should remain singular at any  value of $t$ \cite{Whi55,AGV}.  
To illustrate this point let us consider classical stable mappings at $n=2$ and $n=3$, i.e. folds, cusps and swallow tail \cite{Whi55,AGV} as the initial 
($t=0$)  mapping (\ref{ininhodomap}). Then for the fold at $n=2$ the mapping (\ref{nhodomap}) is given by
\begin{equation}
\begin{split}
x_1=&u_1^2+u_1 t\, , \\
x_2=&u_2+u_2 t\, , 
\end{split}
\label{2Dfold}
\end{equation}
 that corresponds to the solution
 \begin{equation}
 u_1=\frac{1}{2} \left(-t\pm \sqrt{t^2+4 x_1}\right)\, , \qquad u_2=\frac{x_2}{t+1}\, ,
 \end{equation}
of the Euler equation.  The Jacobian of the mapping (\ref{2Dfold}) is
\begin{equation}
J=(1+t)(2u_1+t)\, .
\label{J2Dfold}
\end{equation}
At $t=0$ the mapping (\ref{2Dfold}) is singular on the line $u_1=0$ and for any $t$ it is singular on the line $u_1=-t/2$. At $t=-1$ the mapping degenerates. \par

For the cusp at $n=2$ the mapping (\ref{nhodomap}) is
\begin{equation}
\begin{split}
x_1=&u_1^3+u_1u_2+u_1 t\, , \\
x_2=&u_2+u_2 t\, , 
\end{split}
\label{2Dcusp}
\end{equation}
and the corresponding solution of the Euler equation is
\begin{equation}
u_1= \left( \frac{x_1}{2} + \sqrt{\frac{x_1^2}{4} + \frac{\frac{x_2}{1+t}+t}{27} } \right)^{1/3} 
+  \left( \frac{x_1}{2} - \sqrt{\frac{x_1^2}{4} + \frac{\frac{x_2}{1+t}+t}{27} } \right)^{1/3} \, , \qquad u_2=\frac{x_2}{1+t}\, .
\end{equation}
The Jacobian for this mapping is
\begin{equation}
J=(1+t)(3u_1^2+u_2+t)\, .
\label{J2Dcusp}
\end{equation}
So, the mapping (\ref{2Dcusp}) is singular for any $t$ on the parabolas
\begin{equation}
3u_1^2+u_2+t=0\, .
\end{equation}
In the three-dimensional case the corresponding fold and cusp singularities \cite{AGV} are associated with  the mapping
\begin{equation}
\begin{split}
x_1=&u_1^2+u_1 t\, , \\
x_2=&u_2+u_2 t\, , \\
x_3=&u_3+u_3 t\, , 
\end{split}
\label{3Dfold}
\end{equation}
and
\begin{equation}
\begin{split}
x_1=&u_1^3+u_1u_2+u_1 t\, , \\
x_2=&u_2+u_2 t\, , \\
x_3=&u_3+u_3 t\, , 
\end{split}
\label{3Dcusp}
\end{equation}
Their Jacobian are again given by the formulas (\ref{J2Dfold}) and (\ref{J2Dcusp}). Hence, the mappings (\ref{3Dfold}) and  (\ref{3Dcusp})
are singular for any $t$ on the hypersurfaces in $\mathbb{R}^3$ given by
\begin{equation}
2u_1+t=0\, ,
\end{equation}
and 
\begin{equation}
3u_1^2+u_2+t=0\, .
\end{equation}
For the swallow tail \cite{AGV} the mapping (\ref{nhodomap}) assumes the form
\begin{equation}
\begin{split}
x_1=&u_1^4+u_1u_2^2+u_3 u_1+u_1 t\, , \\
x_2=&u_2+u_2 t\, , \\
x_3=&u_3+u_3 t\, , 
\end{split}
\label{3Dswallowtail}
\end{equation}
The Jacobian of this mapping is 
\begin{equation}
J=(1+t)^2 (4 u_1^3+ 2 u_1 u_2 +u_3+t)\, .
\label{J3Dswallowtail}
\end{equation}
So, the mapping (\ref{3Dswallowtail}) is singular for any $t$ on the hypersurface
\begin{equation}
4 u_1^3 +2 u_1 u_2 +u_3+t=0\, . 
\end{equation}
For $n>3$ one has similar situation.\par

For all these mappings the singularity hypersurface has a single branch defined by the equation
\begin{equation}
J(\uu,t=0)+t=0\, .
\end{equation}
Note that $J=\det M$ where the matrix $M$ is defined by (\ref{Mdef}). Thus, for above mapping the blow-up hypersurfaces (\ref{surcatpoly})
are given by 
\begin{equation}
\det M(\uu,t=0)+t=0\, ,
\end{equation}
and corresponding solutions of the homogeneous Euler equation exhibit blow-up at any time $t$ (for different values of $\uu$). 
\section{Potential case} \label{potential-sec}

Potential flows represent themselves the particular subclass of solutions of $n$-dimensional homogeneous Euler equation for which all branches 
of the blow-up hypersurface are real and consequently any potential solution of (\ref{nBH}) for any dimension $n$ exhibits blow-up.
Indeed, the existence of a potential $\phi$ such that $u_i=\pp{\phi}{x_i}$, $ i=1,\dots,n$  implies that
\begin{equation}
\pp{u_i}{x_k}=\pp{u_k}{x_i}\, ,  \qquad i,k=1,\dots,n\, .
\label{compcondpot}
\end{equation}
Hence, due to the relation (\ref{uder}), the matrix $M^{-1}$ and, consequently, the matrix $M$ (\ref{Mdef}) are symmetric one. Thus, 
\begin{equation}
\pp{f_i}{u_k}=\pp{f_k}{u_i}\, ,  \qquad i,k=1,\dots,n\, ,
\end{equation}
and, so
\begin{equation}
f_i=\pp{\wW}{u_i}\, ,  \qquad i=1,\dots,n\, ,
\label{inipot}
\end{equation}
where $\wW(\uu)$ is some function.
Hence, one has
\begin{equation}
M_{ik}=\frac{\partial^2 \wW}{\partial u_i\partial u_k} + t \delta_{ik} \, ,  \qquad i,k=1,\dots,n\, .
\end{equation}

Thus, in this case the equation (\ref{catMcond}) or (\ref{surcatpoly}) defining the blow-up hypersurface is the characteristic equation for the symmetric matrix 
$\frac{\partial^2 \wW}{\partial u_i\partial u_k}$ and values ${t_0}_i$ in (\ref{tbranches}) coincide (up to a sign) with the eigenvalues of the matrix 
$\frac{\partial^2 \wW}{\partial u_i\partial u_k}$. Consequently, due to the standard properties of the eigenvalues of a real-valued symmetric matrix all branches 
(\ref{tbranches})  are real. It is emphasized that this property of the potential flows is valid for any dimension $n$. Of course, for particular potential flows
(particular functions $\wW$) some branches (\ref{tbranches})  may coalesce. Anyway there is always at least one real branch. This implies  that
any potential solution of the homogeneous Euler equation (\ref{nBH}) in any dimension exhibits blow-up (GC occurs only if the 
critical time is positive). \par

We note that for potential flows the hodograph equation (\ref{nhodo}) represent themselves the equations for the critical points
\begin{equation}
\pp{W}{u_i}=0\, ,\qquad i=1,\dots, n\, ,
\end{equation}
for the function (see also \cite{KO21})
\begin{equation}
W=-\sum_{i=1}^n u_i x_i + \frac{t}{2} \sum_{i=1}^n u_i^2+\wW(\uu)\, .
\label{potsymm}
\end{equation}
At the same time the $\mathbb{R}^n \to \mathbb{R}^n$ mapping (\ref{nhodomap}) is of the form
\begin{equation}
x_i=\pp{W^*}{u_i}\, ,\qquad i=1,\dots, n\, ,
\label{xWmap}
\end{equation}
where
  \begin{equation}
W^*= \frac{t}{2} \sum_{i=1}^n u_i^2+\wW(\uu)\, .
\label{potstarsymm}
\end{equation}
So, for the potential solutions of the homogeneous Euler equation (\ref{nBH}), the associated mappings are the gradient mappings  (\ref{xWmap}). \par

We note also the  well known fact that for potential flows  the homogeneous Euler equation (\ref{nBH}) is equivalent, when $u_i$ is asymptotically zero,  to a Hamilton-Jacobi equation
\begin{equation}
\pp{\varphi}{t}+ \frac{1}{2} \sum_{i=1}^n \left(\pp{\varphi}{x_i}\right)^2=0\, , \qquad \mathrm{with} \qquad \uu \equiv \nabla \varphi\, .
\label{potnBH}
\end{equation}
{It is noted that most solutions of the homogeneous Euler equation discussed in \cite{SZ89,Zel70} (see also \cite{BK07}) correspond to the potential flows.}
\section{Two-dimensional case} \label{2Dcase-sec}
In order to simplify notation we denote $x_1=x$, $x_2=y$, $u_1=u$, $u_2=v$, $f_1=f$, $f_2=g$.  So, the hodograph equations are
\begin{equation}
x=ut+f(u,v)\, , \qquad  y=vt+g(u,v)\, .
\end{equation}
The blow-up hypersurface (\ref{surcatpoly}) is of the form
\begin{equation}
t^2+(f_u+g_v)t+f_ug_v-f_vg_u=0\, .
\label{surcat2}
\end{equation}
Two roots of the equation (\ref{surcat2}) are given by
\begin{equation}
\begin{split}
t_{\pm}(u,v)&= \frac{1}{2}\left( -(f_u+g_v) \pm \sqrt{(f_u+g_v)^2-4(f_ug_v-g_u f_v) }\right) \\
&= \frac{1}{2}\left( -(f_u+g_v) \pm \sqrt{(f_u-g_v)^2+4g_u f_v}\right)  \, .
\end{split}
\end{equation}
So, the blow-up occurs if 
\begin{equation}
\Delta \equiv  (f_u+g_v)^2 - 4 J_0=(f_u-g_v)^2+4g_u f_v \geq 0\, ,
\label{condcat2}
\end{equation}
in the domain $\mathcal{D}_\uu$ where $J_0=f_ug_v-g_u f_v$. In the case $\Delta=0$, i.e. when the functions $f$ and $g$ obey the PDE
\begin{equation}
(f_u-g_v)^2+4g_u f_v = 0\, ,
\label{degcondcat2}
\end{equation}
 there is a single branch and at $f_u+g_v>0$ one has blow-up at $t=-\frac{1}{2} (f_u+g_v)<0$. In the opposite case $f_u+g_v<0$, the corresponding 
 solution have GC at $t={\min}_{u,v} \left[ -\frac{1}{2}(f_u+g_v) \right]>0$.  \par
 
 If $\Delta>0$  in the domain $\mathcal{D}_\uu$ there are three different cases.  
 \begin{enumerate}
 \item[1st case]: $f_u+g_v>0$, $J_0>0$. The corresponding solution exhibit two blow-ups at negatives
 times on $t_+$ and $t_-$.
 \item[2nd case]: $f_u+g_v>0$, $J_0>0$.  One has GCs on the branches $t_+$ and $t_-$.
 \item[3rd case]: $J_0<0$. One has blow-up on the branch $t_-<0$ and GC at times $t_+>0$.
 \end{enumerate}
If $J_0=0$ one has blow-ups on the branches $t_+=0$ and $t_-=-{f_u+g_v}$.\par

 It is easy to see that for potential flows ($u=\varphi_x$, $v=\varphi_y$) the condition $\Delta>0$ (\ref{condcat2})
is always satisfied. Indeed, since in such a case (see  (\ref{inipot}))
\begin{equation}
f=\pp{\wW}{u}\, , \qquad g=\pp{\wW}{v}\, ,
\label{2inipot}
\end{equation}
then 
\begin{equation}
\Delta=(\wW_{uu}-\wW_{vv})^2+4(\wW_{uv})^2 > 0\, ,
\end{equation}
for any  functions $\wW$  (except the trivial  case with $\wW=A (u^2+v^2)+Bu+Cu+D$ for which $\Delta=0$). \par

The  case when $\Delta$ is negative corresponds to the solution of $2D$ homogeneous Euler equation free of blow-ups.
In order to analyse such and other situations in $2D$ case, it is convenient to rewrite the $2D$ Euler equation and the corresponding hodograph 
equations in complex variables. For this purpose we introduce the notations
\begin{equation}
z=x+iy\, , \qquad V=u+iv\, , \qquad F=f+ig\, .
\end{equation}
In these variables the $2D$ homogeneous Euler equation assumes the form
\begin{equation}
V_t+VV_z +\ccV V_\ccz=0\, ,
\label{2BHc}
\end{equation}
  and hodograph equations become
  \begin{equation}
  z=Vt+F(V,\ccV)\, .
  \label{2Dchodo}
  \end{equation}
Equation (\ref{2Dchodo}) implies that 
\begin{equation}
\begin{split}
V_z=\frac{\cc{F}_\ccV+t}{\det M}\, , \qquad
V_\ccz=-\frac{{F}_\ccV}{\det M}\, , \qquad
V_t=\frac{-V(\cc{F}_\ccV+t )+\ccV \ {F}_\ccV}{\det M}\, ,  
\end{split}
\label{solcc2D}
\end{equation}
where
\begin{equation}
\det M= (F_V+t)(\cc{F}_\ccV+t)-F_\ccV \cc{F}_V\, .
\label{detMcondcc2D}
\end{equation}
So two roots of $\det M=0$ composing the blow-up surface are given by
\begin{equation}
t_{\pm}(u,v)=\frac{1}{2} \left( - (F_V+\cc{F}_{\ccV}) \pm \sqrt{(F_V-\cc{F}_\ccV)^2+4 F_\ccV \cc{F}_V} \right)= - \Re(F_V) \pm \sqrt{-(\Im(F_V))^2+|F_\ccV|^2} \, .
\label{c2Dtbranches}
\end{equation}
Now let us consider two particular subclasses of initial data, i.e. functions $F$. First class is given by analytic functions $F$, i.e. $F_\ccV=0$. In such case 
$V_\ccz=0$ and the Euler equation (\ref{2BHc}) becomes
\begin{equation}
V_t+VV_z=0\, .
\label{2BHcanal}
\end{equation}
This complex Burgers-Hopf equation has been considered earlier in the papers \cite{KSZ94,KZ14,ZK18} within the study of potential two-dimensional flows in particular hydrodynamical problems.\par

The reduction (\ref{2BHcanal}) of the 2D Euler equation (\ref{2BHc}) has one particular property. In the domain where $F_\ccV=0$ (and hence $V_\ccz=0$)
the expression in the square root is always negative ($ (F_V-\cc{F}_\ccV)^2=-(\Im(F_V))^2<0$). Thus, these solutions of the equation (\ref{2BHcanal})
are blow-up free.  Note that the singularities for equation (\ref{2BHcanal}) discussed in \cite{KZ14,ZK18} correspond to the singularities of $F(V)$ (poles, etc.)
in nonphysical domain.\par

Another interesting reduction of the equation (\ref{2BHc})  is given by  the mapping $(z,\ccz)\to(V,\ccV)$, defined by the 
Beltrami equation \cite{Ahl}
\begin{equation}
V_\ccz=\mu(z,\ccz,t) V_z\, ,
\label{Beltrami}
\end{equation}
where $\mu$ is the so-called complex dilation. In the case $|\mu|<1$ the mapping is quasi-conformal \cite{Ahl}.

Under this constraint the equation (\ref{2BHc}) becomes
\begin{equation}
V_t+(V+\mu \ccV)V_z=0\, .
\label{2BH-Beltrami}
\end{equation}
The consistency of the constraint (\ref{Beltrami}) with the equation (\ref{2BH-Beltrami}) requires that $\mu$ obeys the following equation
\begin{equation}
\mu_t V_z- \mu ((V+\mu \ccV)V_z)_z+((V+\mu \ccV)V_z)_\ccz=0\, .
\label{compmu}
\end{equation}
The first two formulae (\ref{solcc2D}) imply that the constraint (\ref{Beltrami}) is equivalent to the following one
\begin{equation}
F_\ccV=-\mu(\cc{F}_\ccV+t)\, ,
\label{equivBel}
\end{equation}
or
\begin{equation}
z_\ccV=-\mu \ccz_\ccV\, .
\label{equivBel-hodo}
\end{equation}
Equation defining the blow-up surface $\det M=0$ (see \ref{detMcondcc2D}) in this case assumes the form 
\begin{equation}
F_\ccV \cc{F}_V (|\mu|^2-1)=0\, .
\end{equation}
Thus, solutions of the equation (\ref{2BH-Beltrami}) exhibit blow-ups only in the case $|\mu|=1$. It is exactly the case of singularity
of quasi conformal mapping defined by the constraint (\ref{Beltrami}) (see e.g. \cite{Ahl}). \par

At $|\mu|<1$ the quasi-conformal mapping (\ref{Beltrami}) and solutions of the equation (\ref{2BHc}) are free of singularities. 
\FloatBarrier
\section{Two dimensional case: examples} \label{2Dexa-sec}
In this section we will present some illustrative examples. 
\subsection{First example}
\begin{figure}[h!]
\begin{center}
\includegraphics[width=.3 \textwidth]{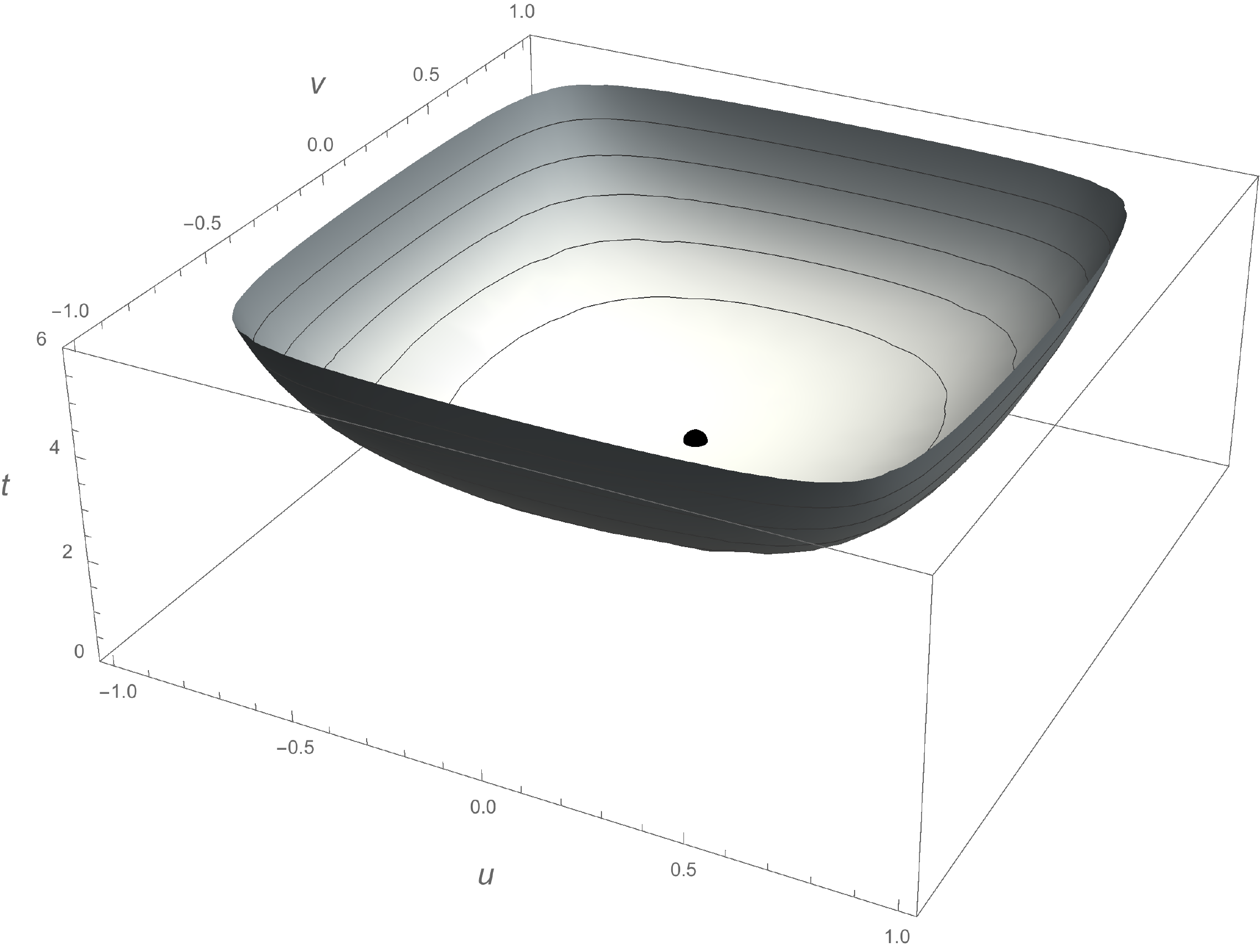}
\caption{Evolution of the multivalued region with initial data given in (\ref{index1}). The black dot identifies the catastrophe time.
The folding region is the region where the velocity $\uu$ is multivalued: this corresponds to the region enclosed by the line of blow-up. }
\label{tc3Dplot-fig}
\end{center}
\end{figure}

We begin with the simple example corresponding to  the initial data
\begin{equation}
u_0=\tanh(x+2y)\, , \qquad  v_0=\tanh(x+y)\, .
\label{index1}
\end{equation}
So, the hodograph equations are
\begin{equation}
x-ut=-\, \atanh(u)+2 {\atanh(v)}\, , \qquad y-vt=\atanh(u)-\atanh(v)\, ,
\label{solex1}
\end{equation}
where the domain $\mathcal{D}_\uu$  is the square $-1 \leq u \leq 1$, $-1 \leq v \leq 1$.
The corresponding Jacobian  is
\begin{equation}
\left( \begin{array}{cc}
f_{u}&g_{v}\\g_{u}&g_{v}
\end{array}
\right)= 
\left(
\begin{array}{cc}
 -\frac{1}{1-u^2} & \frac{2}{1-v^2} \\
 \frac{1}{1-u^2} & -\frac{1}{1-v^2} \\
\end{array}
\right)\, .
\end{equation}
and the blow-up surface is given by the equation 
\begin{equation}
t^2-\frac{2-u^2-v^2}{(1-u^2)(1-v^2)} t -\frac{1}{(1-u^2)(1-v^2)}=0\, .
\end{equation}
It has two branches $t=t_+ \cup t_-$ 
\begin{equation}
t_\pm (u,v)= \frac{2-u^2-v^2\pm\sqrt{(2-u^2-v^2)^2+4(1-u^2)(1-v^2)}}{2(1-u^2)(1-v^2)} 
\, .
\end{equation}
It is easy to see that for the $(+)$ branch $t>0$ always and  for the $(-)$ branch $t<0$. 
We report in figure \ref{tc3Dplot-fig}  the plot of $(+)$ branch. Thus, one has GC on the $(+)$ branch at $t_c=1+\sqrt{2}$ at $u_c=v_c=0$ (and a blow-up on the $(-)$
branch for $t<1-\sqrt{2}$ starting from $u_c=v_c=0$). 
Finally from the hodograph solution (\ref{solex1}) we get the catastrophe position $x_c=y_c=0$.\par

In the case (\ref{solex1}) and at the point $t_c=1+\sqrt{2}$, the matrices $M(0)$  and $\widetilde{M}(0)$ are of the form 
\begin{equation}
M(0)=\left( \begin{array}{cc}
\sqrt{2} & 2 \\ 1 & \sqrt{2}
\end{array}
 \right)\, , \qquad
 \widetilde{M}(0)=\left( \begin{array}{cc}
\sqrt{2} & -2 \\ -1 & \sqrt{2}
\end{array}
 \right)\, .
\label{M0solex1}
\end{equation}
Hence, the vectors $\vec{\widetilde{R}}$, $\vec{\widetilde{L}}$, $\vec{{R}}$, and $\vec{{L}}$, defined by (\ref{reigen}), (\ref{leigen}),  (\ref{reM}), and (\ref{leM}) are
\begin{equation}
\vec{\widetilde{R}}=(\sqrt{2},1)a \, , \quad \vec{\widetilde{L}}=(1,\sqrt{2})b \, , \quad
\vec{{R}}=(-\sqrt{2},1)c \, , \quad \vec{{L}}=(1,-\sqrt{2})d \, ,
\end{equation}
where $a,b,c,d$ are arbitrary real constants. So, at the of GC described above, the following combinations of blow-up derivatives remain finite {at time $t_c=1+\sqrt{2}$}
\begin{equation}
\begin{split}
&\sqrt{2} \pp{u_i}{x}+\pp{u_i}{y} \sim O(1)\, , \qquad i=1,2\, , \\
 &\pp{u_1}{x}+\sqrt{2}\pp{u_2}{x} \sim O(1)\, , \quad
 \pp{u_1}{y}+\sqrt{2}\pp{u_2}{y} \sim O(1)\, . 
\end{split}
 \end{equation}
On the other hand, the matrices $\frac{\partial^2 f_i}{\partial u_k \partial u_l}$ $i=1,2$ are identically zero in the catastrophe point $(u,v)=(u_c,v_c)=(0,0)$.
So, the second order term in the r.h.s. of the formula (\ref{xdevu}) vanishes and consequently the variations of $x_i$ are cubic polynomials in $\delta u_i$. 
This is a consequence of the fact that the GC point $t_c=1+\sqrt{2}$, being the first  blow-up point, is non-generic.

\subsection{$\mathbf{1+\epsilon}$ dimensional  case}

As the second example we consider the situation when initial data (and so functions $f$ and $g$) depend on parameter. Namely, let
 \begin{equation}
 u(\ux,0)= \tanh(x+\epsilon y)\, , \qquad  v(\ux,0)=  \tanh(\epsilon x+ y)\, , \qquad \epsilon \geq 0\, .
 \label{1+eini}
 \end{equation}
 At $\epsilon =0$ one has the decomposition into two BH equation with no GC for both fields $u$ and $v$. At $\epsilon=1$ the problem degenerates to the 
 reduction $u=v$.
 For the initial data (\ref{1+eini}) the hodograph equations are
\begin{equation}
x-ut=\frac{\epsilon  \atanh(v)-\atanh(u)}{\epsilon ^2-1}\, , \qquad y-vt=\frac{\epsilon  \atanh(u)-\atanh(v)}{\epsilon ^2-1}\, .
\label{sol1+e}
\end{equation}
and the domain $\mathcal{D}_\uu$ is the square $-1\leq u\leq 1$, $-1\leq v\leq 1$.
The blow-up surface is defined by the equation
\begin{equation}
t^2-
\frac{2-u^2-v^2}{\left(1-u^2\right) \left(1-v^2\right) \left(\epsilon
   ^2-1\right)}t
   -\frac{1}{\left(1-u^2\right) \left(1-v^2\right) \left(\epsilon
   ^2-1\right)}=0
\label{par1+e}
\end{equation}

%
Two branches $t_\pm$ of the surface $t=t_+ \cup t_-$ defined by (\ref{par1+e}) are given by
\begin{equation}
t_\pm=\frac{ 2-u^2-v^2 \pm \sqrt{\left(2-u^2-v^2\right)^2+4 \left(1-u^2\right) \left(1-v^2\right)( \epsilon ^2-1)}}
{2\left(1-u^2\right) \left(1-v^2\right) \left(\epsilon^2-1\right)}\, .
\end{equation}
It is easy to see that for the $(-)$  branch $t$ is negative for all values of $\epsilon$. Instead, for the $(+)$ branch $t>0$ at $\epsilon>1$ and 
$t < 0$ at $0 < \epsilon < 1$.   \par

The catastrophe time $t_c$ coincides with the minimum  $t_\mathrm{min}$ of the $(+)$ branch.
Evaluating the $t_\mathrm{min}$ for the $(+)$ branch one finds that it corresponds to $u_c=v_c=0$ and 
\begin{equation}
t_\mathrm{min}=t_c=\frac{1}{\epsilon-1}\, .
\end{equation}
Thus, the solutions of the 2d-homogeneous Euler equation with the initial data (\ref{1+eini}) and ``small'' coupling of dimensions ($0<\epsilon <1$) do not exhibit GC as in 
the decoupled case ($\epsilon=0$). For $\epsilon>1$ the time of the first appearance of GC decreases with increasing of the coupling $\epsilon$. 

Evaluating $t$ for the $(-)$ branch at $u=v=0$ one gets
\begin{equation}
t_\mathrm{-,max}= -\frac{1}{\epsilon+1}\, .
\end{equation}
So, for $0<\epsilon<1$ 
\begin{equation}
t_\mathrm{-,max}>t_\mathrm{min}\, .
\end{equation}
Thus,  at $0<\epsilon<1$ the solution (\ref{sol1+e}) exhibit blow-ups at negative $t$ with maximum given by $t_\mathrm{-,max}$.
\subsection{Generic 2d case}
Let us consider the case which initial data which resemble standard initial data in the one-dimensional case, namely,
\begin{equation}
u(x,0) = u_0 (1-\tanh(\alpha x + \beta y))\, , \qquad
v(x,0) = v_0 (1-\tanh(\gamma x + \delta y))\, ,
\end{equation}
where $u_0,v_0,\alpha,\beta,\gamma,\delta$ are constants, $u_0,v_0>0$ and the domain $\mathcal{D}_\uu$ is the rectangle $0\leq u \leq 2u_0$,
$0\leq v \leq 2v_0$. The corresponding hodograph equations  are 
\begin{equation}
\begin{split}
x-ut=& \frac{1}{\Delta} \left(\delta \atanh \left( 1-\frac{u}{u_0}  \right)-\beta  \atanh \left( 1-\frac{u}{u_0}     \right) \right) \, , \\
y-vt= &\frac{1}{\Delta} \left( -\gamma \atanh \left( 1-\frac{u}{u_0} \right) +\alpha  \atanh \left( 1-\frac{u}{u_0}  \right) \right) \, , 
\end{split}
\label{2dexe-mixsol}
\end{equation}
where $\Delta=\alpha \delta-\beta \gamma$. One has 
\begin{equation}
J=\frac{1}{\Delta}  \left(
\begin{array}{cc}
 -\frac{\delta  u_0}{u(2  u_0-u)} & \frac{\beta  {v_0}}{v(2  v_0-v)} \\
 \frac{\gamma  {u_0}}{u(2  u_0-u)} & -\frac{\alpha  {v_0}}{v(2  v_0-v)} \\
\end{array}
\right)
\end{equation}
and 
\begin{equation}
\begin{split}
T\equiv & \tr{J}=-\frac{1}{\Delta} \frac{\delta  u_0 v(2  v_0-v) +\alpha v_0 u(2  u_0-u)}{uv(2  u_0-u)(2  v_0-v)}\, , \\
D\equiv & \det{J}= \frac{1}{\Delta} \frac{u_0v_0}{uv(2  u_0-u)(2  v_0-v)}\, .
\end{split}
\end{equation}
So the blow-up surface is $t=t_+ \cup t_-$ with the two branches given by
\begin{equation}
t_\pm = \frac{T\pm \sqrt{T^2-4D}}{2}\, .
\end{equation}
Critical values of $t$, given by ${t_\pm}_u={t_\pm}_v=0$, are  reached where $u=u_0$ and $v=v_0$  corresponding to
\begin{equation}
{t_\pm}_c=\frac{\alpha u_0 + \delta v_0 \pm \sqrt{(\alpha u_0 + \delta v_0)^2-4 \Delta u_0v_0}}{2 \Delta u_0v_0}\, .
\label{tc-abcd2D}
\end{equation}
The value ${t_\pm}_c$ is real when
\begin{equation}
\Delta<\frac{(\alpha u_0 + \delta v_0)^2 }{4  u_0v_0}\, .
\end{equation}
Let us now consider some particular cases. 

In the decoupled case when $\beta=\gamma=0$ then $\Delta=\alpha \delta$ and 
\begin{equation}
{t_+}_c=\frac{1}{\delta v_0}\, , \qquad  {t_-}_c=\frac{1}{\alpha u_0}\, .
\end{equation}
So one gets, for the $t_\mathrm{min}$ for the GC
\begin{equation}
\begin{split}
{t_-}_\mathrm{min}=\frac{1}{\alpha u_0}\, , \qquad \mathrm{for\,  the\,  field\, } u\, , \\
{t_+}_\mathrm{min}=\frac{1}{\delta v_0}\, , \qquad \mathrm{for\,  the\,  field\, } v \, .
\end{split}
\label{2BHabcd}
\end{equation}
Therefore the appearance of  GC is resumed in the following table
\begin{equation}
\begin{array}{|c|c|c|}
\hline
\alpha&\delta& \mathrm{fields \, with \, GC} \\
\hline
>0&>0& u,v \\
\hline
>0&<0& u \\
\hline
<0&>0& v \\
\hline
<0&<0& \mathrm{none} \\
\hline
\end{array}
\end{equation}
This table reproducts well known situation for GC of BH equation. The same results are obtained in the semi-decomposed case $\beta \neq 0$, $\gamma=0$ or
$\beta = 0$, $\gamma \neq 0$. \par

The quasi-decomposed case  corresponds to $\beta=\gamma=\epsilon>0 $ where $\epsilon=o(\alpha)=o(\beta)$.
The behavior for small $\epsilon$ of the blow-up time (\ref{tc-abcd2D}) depends critically on values of  $\alpha u_0 - \delta v_0$:
 If $\alpha u_0 - \delta v_0 \neq 0$ the correction to the decouple case (\ref{2BHabcd}) is quadratic in $\epsilon$
 \begin{equation}
\begin{split}
{t_+}_c=&\frac{1}{\delta  {v_0}} + \frac{{u_0}}{ \delta ^2  {v_0}(\alpha u_0 -\delta  {v_0})}   \epsilon ^2+  O(\epsilon^3) \, ,\\
{t_-}_c=&\frac{1}{\alpha  {u_0}}- \frac{{v_0}}{ \alpha^2  {u_0}(\alpha u_0 -\delta  {v_0})}   \epsilon ^2+ O(\epsilon^3) \, .
\end{split}
\end{equation}
Otherwise, if $\alpha u_0 - \delta v_0 = 0$ the corrective term is linear in $\epsilon$
\begin{equation}
\begin{split}
{t_+}_c=&\frac{1}{\delta  {v_0}}+\frac{1 }{\alpha\delta \sqrt{u_0v_0}}\epsilon+ O(\epsilon^2) \, ,\\
{t_-}_c=&\frac{1}{\alpha  {u_0}}- \frac{1 }{\alpha\delta \sqrt{u_0v_0}}\epsilon+ O(\epsilon^2) \, .
\end{split}
\end{equation}
\subsection{Direct case example and qualitative comparison with numerics}
When the local inversion of initial data is computationally complicated, the mathematical characterization of the blow-ups and catastrophes could be  done
 using a direct characteristic-like method  depicted in appendix \ref{char-app}. \par
 
\begin{figure}[h!]
\begin{center}
\begin{tabular}{cc}
\includegraphics[width=.3 \textwidth]{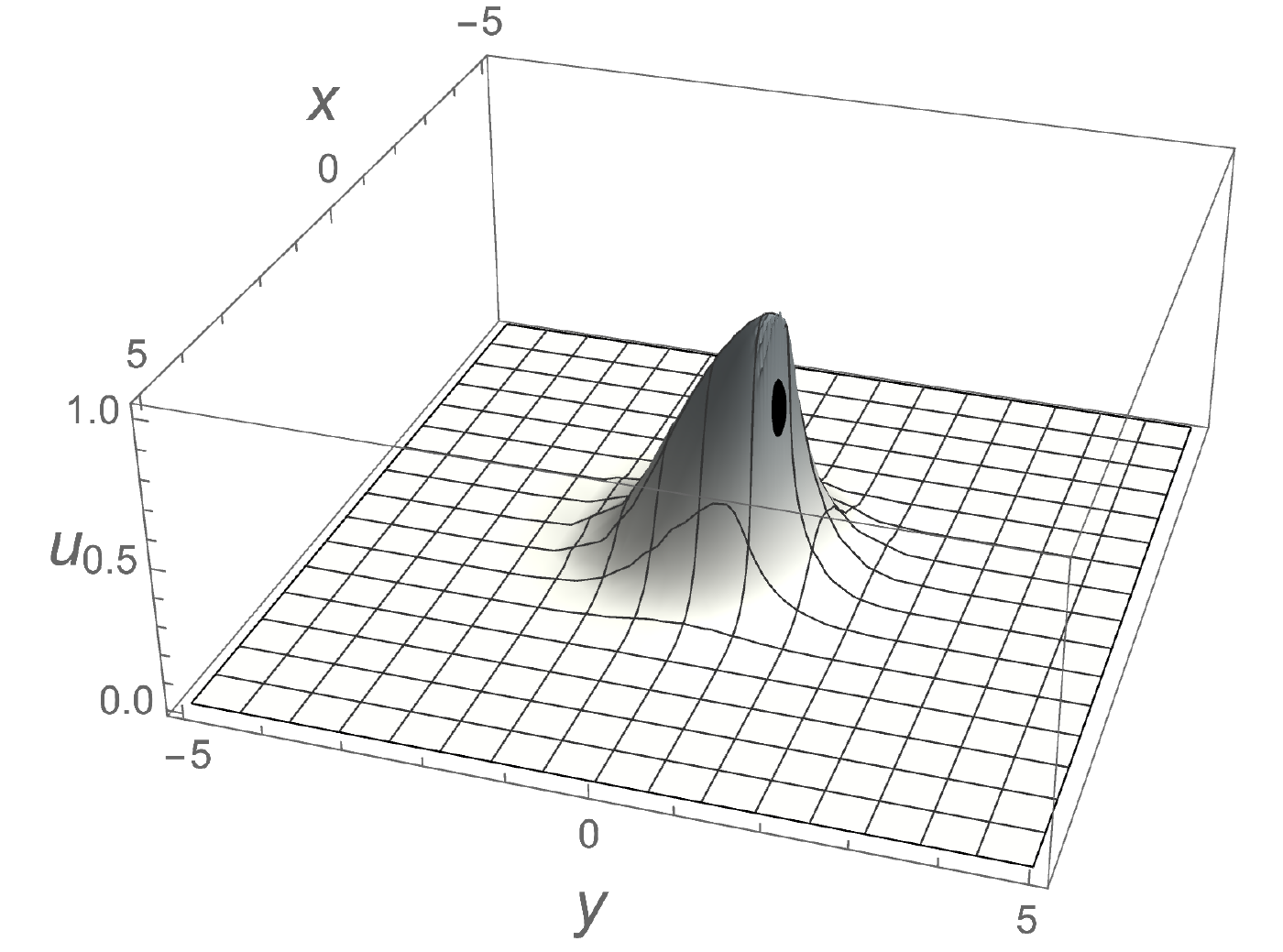} & \includegraphics[width=.3 \textwidth]{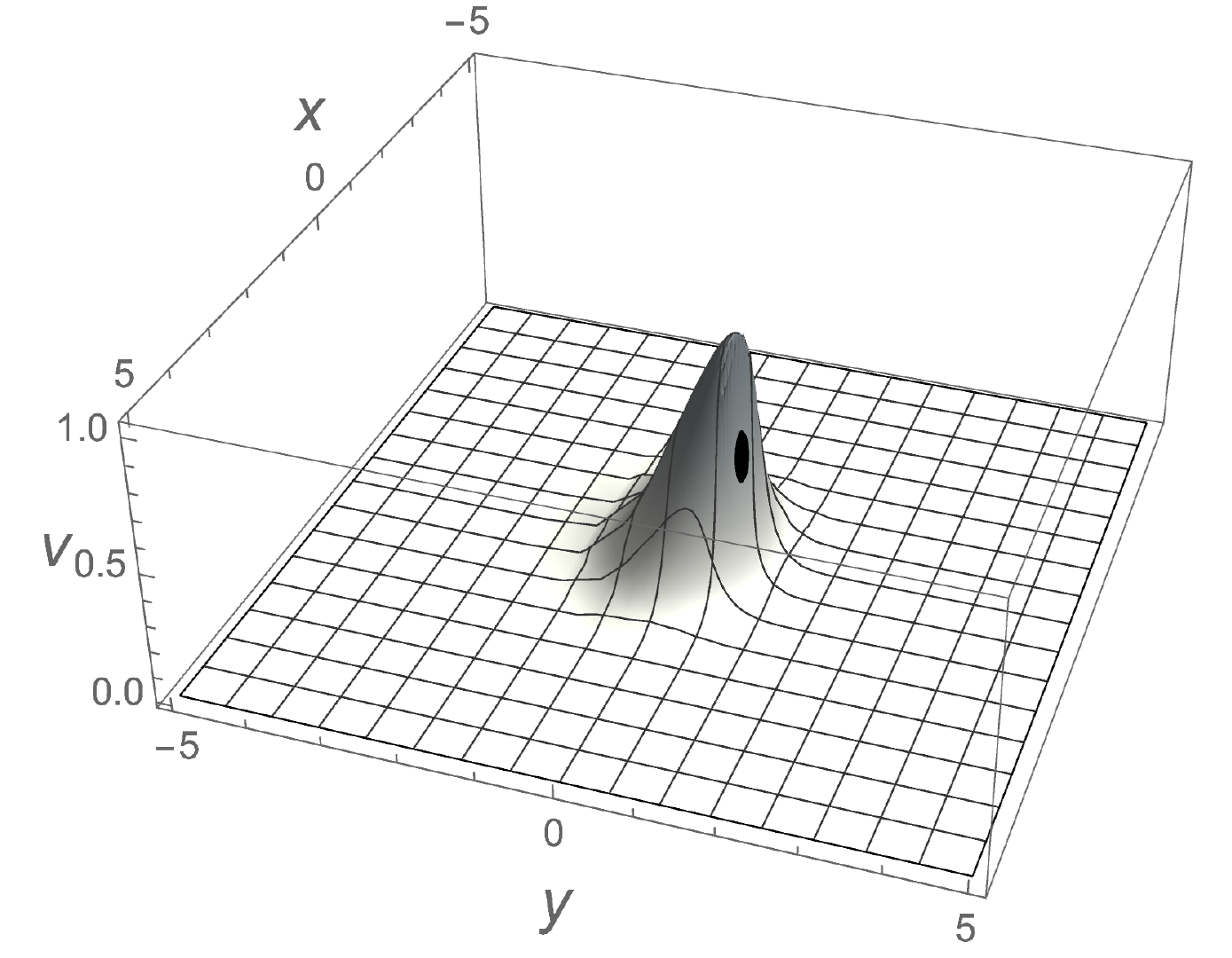} 
\end{tabular}
\caption{Evolution for initial data given in (\ref{exe-exp2D}) at catastrophe time ($t= 0.728$): the black dot is the catastrophe point.}
\label{Exp2D-fig}
\end{center}
\end{figure}

 Let us consider the initial data
\begin{equation}
u(x,0)=e^{-x^2-y^2}\, , \qquad v(x,0)=e^{-x^2-2y^2}\, .
\label{exe-exp2D}
\end{equation}
In this case the local inversion of (\ref{exe-exp2D}) is possible but it would require the study of many local inverses: we will use the direct method.
The Jacobian of the initial data  (\ref{exe-exp2D}) is
\begin{equation}
\left(
\begin{array}{cc}
 -2 x e^{-x^2-y^2} & -2 y e^{-x^2-y^2} \\
 -2 x e^{-x^2-2 y^2} & -4 y e^{-x^2-2 y^2} \\
\end{array}
\right) .
\end{equation}
The opposite inverse eigenvalues of such matrix  give the blow-up region $t=t_+ \cup t_-$ where
\begin{equation}
t_\pm(x_0,y_0)=\frac{e^{x_0^2+2 y_0^2}}{x_0 e^{y_0^2}+2 y_0 \pm \sqrt{x_0^2 e^{2 y_0^2}+4 y_0^2}}
\label{exe-bureg}
\end{equation}
where $\ux_0$ is a suitable parameter related to characteristics (see appendix \ref{char-app}).
Using Mathematica one can compute the minimum of the functions (\ref{exe-bureg}) and the other chatastrophe parameters obtaining
\begin{equation}
t_c=0.7281359\, , \qquad u_c= 0.705897 \, , \qquad v_c= 0.572292 \, , \qquad x_c= 0.886099\, , \qquad y_c= 0.874767 \,  .
\end{equation}
In figure \ref{Exp2D-fig} we compare the numerical evolution of the initial data (\ref{exe-exp2D}) with the catastrophe point computed by the theory.

\section{Dynamics of mappings. Examples in two-dimensional case} \label{mapexe-sec}
\begin{figure}[h!]
\begin{center}
\includegraphics[width=.2 \textwidth]{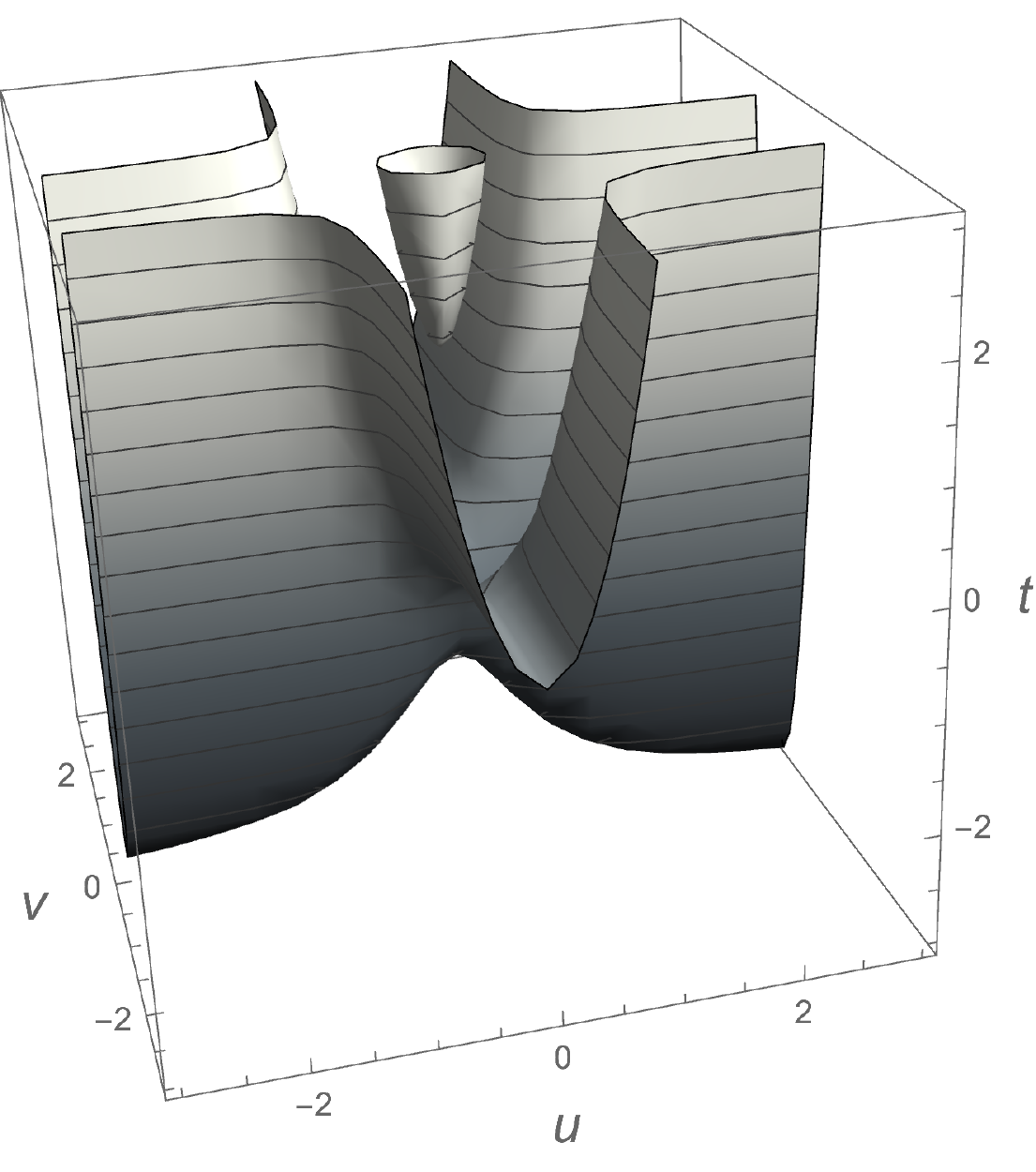} 
\caption{Plot of the the blow-up region $t=t(u,v)$ implicitly defined by (\ref{bureg-exe}). }
\label{mapexe-fig}
\end{center}
\end{figure}

Let us consider the hodograph equation  from the point of view of the mappings.
The first example is
\begin{equation}
x=u t- 2 u^3 - 2 v - v^3\, , \qquad  y= v t- u - 5 v^3 - 3 u^3\, .
\label{map-exe}
\end{equation}
The Jacobian of the map (\ref{map-exe}) looks like
\begin{equation}
M=\left(
\begin{array}{cc}
 t-6 u^2 & -3 v^2-2 \\
 -9 u^2-1 & t-15 v^2 \\
\end{array}
\right)
\end{equation}
and its determinant, giving the blow-up region, is 
\begin{equation}
J=\det(M)=t^2+ \left(-6 u^2-15 v^2\right) t +63 u^2 v^2-18 u^2-3 v^2-2\, .
\label{bureg-exe}
\end{equation}
At $t=0$ the mapping (\ref{map-exe}) is singular on the curve
\begin{equation}
v=\pm\sqrt{\frac{18 u^2 +2}{63 u^2-3}}.
\end{equation}
In the figure  \ref{mapexe-fig}, both the branches of the blow-up region (\ref{bureg-exe}) are plotted. The positive minimum identifies also a catastrophe point.
\par
Now let us consider the family of mappings on the whole plane $(u,v)\to(x,y)$
\begin{equation}
\begin{split}
x=tu-\frac{1}{3}u^3+\frac{2}{3}v^3-u+2v\, , \\
y=tv+\frac{1}{3}u^3-\frac{1}{3}v^3+u-v\, .
\label{exe-map-3}
\end{split}
\end{equation}
It is singular on the curve given by the equation
\begin{equation}
t^2-t \left(2+u^2+v^2\right)-\left(1+u^2\right) \left(1+v^2\right)=0
\label{sing-map-3}
\end{equation}
So, at $t=0$, the mapping (\ref{exe-map-3}) is free of singularities ($\left(1+u^2\right) \left(1+v^2\right) \neq 0$). The curve (\ref{sing-map-3})  defines a surface 
$t=t_+ \cup t_-$ where 
\begin{equation}
t_\pm (u,v)= \frac{1}{2} \left(2+u^2+v^2\pm \sqrt{u^4+v^4+6 u^2 v^2+8 u^2+8 v^2+8}\right)\, .
\end{equation}
The $(+)$-branch is the surface in $\mathbb{R}^3$ with the coordinates  $(t,u,v)$ with the minimum ${t_+}_\mathrm{min}=1+\sqrt{2}$ at the point $u_c=v_c=0$.
The $(-)$-branch is the surface in $\mathbb{R}^3$ with the coordinates  $(t,u,v)$ with the maximum ${t_-}_\mathrm{min}=1-\sqrt{2}$ at the point $u_c=v_c=0$.
 So, the dynamics of the mapping (\ref{exe-map-3}) is rather specific: it has a singularity at $t<1-\sqrt{2}$ on the $(-)$-branch. It has no singularity on the interval
 $1-\sqrt{2}<t<1+\sqrt{2}$ and it becomes again singular for $t>1+\sqrt{2}$ on the $(+)$-branch.\par
 As the third example we consider the mapping
\begin{equation}
\begin{split}
x&=tu+\frac{1}{3}u^3+\frac{2}{3}uv^2-2v\, , \\
y&=tv+\frac{1}{3}v^3-\frac{1}{3}u^2v+u\, .
\end{split}
\label{exe-map-3+}
\end{equation}
The Jacobian of this mapping is  
\begin{equation}
J=t^2+\frac{1}{3} t \left(4 u^2+5 v^2\right)+\frac{1}{3} \left(u^2 v^2+u^4+2 v^4+6\right)\, .
\end{equation}
So, the mapping (\ref{exe-map-3+}) is obviously nonsingular at $t=0$. 
Two branches of the singular surface $t=t_+ \cup t_-$ are given by
\begin{equation}
t_\pm(u,v)=\frac{1}{6} \left(-4 u^2-5 v^2 \pm \sqrt{(4 u^2+5 v^2)^2-12 \left(u^2 v^2+u^4+2 v^4+6\right)}\right)\, .
\end{equation}
So, for both branches $t_\pm<0$.
Thus, the mapping (\ref{exe-map-3+}) has no singularities for $t \geq 0$. Hence, the possible singularities   of such mapping  for $t<0$ disappear at $t\geq 0$.
\section{Three-dimensional examples} \label{3Dexe-sec}
Let us now consider a simple, yet not trivial, example of a 3D case. We rename $\uu$ components as $\uu=(u,v,w)$.
The initial data 
\begin{equation}
u(\ux,0)=\frac{1-\tanh(y)}{2}\, , \qquad v(\ux,0)=\frac{1-\tanh(z)}{2}\, , \qquad w(\ux,0)=\frac{1-\tanh(x)}{2}\, ,
\label{3Dexe-simple}
\end{equation}
lead to the hodograph equations
\begin{equation}
x-ut=\atanh(1-2w)\, , \qquad  y-vt=\atanh(1-2u)\, , \qquad  z-wt=\atanh(1-2v)\, .
\end{equation}
The associated matrix $M$ (\ref{Mdef}) is 
\begin{equation}
M=\left(
\begin{array}{ccc}
 t & 0 & -\frac{2}{1-(1-2 w)^2} \\
 -\frac{2}{1-(1-2 u)^2} & t & 0 \\
 0 & -\frac{2}{1-(1-2 v)^2} & t \\
\end{array}
\right)\, .
\end{equation}
The real solution of the equation $\det (M)=0$ is
\begin{equation}
t_1=\frac{2}{(\left(1-(1-2 u)^2\right) \left(1-(1-2 v)^2\right) \left(1-(1-2 w)^2\right))^{1/3}}\, .
\end{equation}
The branch $t_1(u,v,w)$ admits a positive minimum $t_c \equiv t_1(1/2,1/2,1/2)=2$. Finally, the catastrophe values for the initial data    (\ref{3Dexe-simple}) are
\begin{equation}
t_c=2\, , \qquad u_c=v_c=w_c=1/2\, , \qquad x_c=y_c=z_c=1\, .
\end{equation}
\par

As our last example we consider solution of the 3-dimensional Euler equation with initial data
\begin{equation}
u(\ux,0)=\tanh(x+\epsilon y)\, , \qquad v(\ux,0)=\tanh(y+\epsilon z)\, , \qquad w(\ux,0)=\tanh(z+\epsilon x)\, ,
\label{3Dexe-1}
\end{equation}
where $\epsilon$ is a parameter ($\epsilon \neq 1$). The corresponding hodograph equations are
\begin{equation}
\begin{split}
x-ut=&\frac{1}{1+\epsilon^3}\left( \atanh(u)-\epsilon \atanh(v)+\epsilon^2 \atanh(w)  \right) \\
y-vt=&\frac{1}{1+\epsilon^3}\left( \atanh(v)-\epsilon \atanh(w)+\epsilon^2 \atanh(u)  \right) \\
z-wt=&\frac{1}{1+\epsilon^3}\left( \atanh(w)-\epsilon \atanh(u)+\epsilon^2 \atanh(v)  \right) \, .
\end{split}
\label{exe3D-parsol}
\end{equation}
The domain $\mathcal{D}_\uu$ is given by the cube $-1 \leq u \leq 1$, $-1 \leq v \leq 1$, $-1 \leq w \leq 1$.\par

The blow-up hypersurface is given by the equation
\begin{equation}
\left(t+\frac{1}{1-u^2} \right) \left(t+\frac{1}{1-v^2} \right) \left(t+\frac{1}{1-w^2} \right) +\epsilon^3 t^3=0\, ,
\label{exe3D-par}
\end{equation}
which manifestly shows the ``coupling'' of three one-dimensional BH equations at $\epsilon \neq 0$. Solutions of each of these three BH equations does not
exhibit the GC. In the 3-dimensional case ($\epsilon \neq 0$) situation depends in the value of $\epsilon$. 
The symmetry by permutation on $u,v,w$ of the relation (\ref{exe3D-par}) implies that the minimum $t_c$ of $t$ is reached in $u_c=v_c=w_c$ and by direct computation 
one obtains $u_c=v_c=w_c=0$.
The corresponding value of $t_c$ can be
easily calculated using equation (\ref{exe3D-par}). Indeed one has $(t_c+1)^3+\epsilon^3t_c^3=0$, and hence
\begin{equation}
t_c=-\frac{1}{1+\epsilon}\, .
\end{equation}
Thus the solution (\ref{exe3D-parsol}) exhibits GC for the parameter $\epsilon <-1$.

\section{Conclusion} \label{conclusion-sec}
In this paper we have addressed the problem of the blow-ups and catastrophes  for homogeneous Euler equation (\ref{nBH}).  
The approach is based on the extension of the hodograph method for Burgers-Hopf equation to many dimensions and it allows to associate a mapping  on 
$\mathbb{R}^n$  to every solution. 
A complete classification of the singularities appearing on these mappings has been performed in \cite{KK02}  for the 1D case: 
in the multi-dimensional case, the existence of a hierarchy of  singularities (depending on different classes of initial data)   is a natural question. \par 

Another problem to be addressed is the regularization of the gradient catastrophes described above. In contrast to the multidimensional analogs of the Rankine- Hugoniot condition, shock waves  etc discussed in \cite{Guc75,She02} one can adopt an approach proposed in the one-dimensional case in \cite{KO18}  and develop its multidimensional version. \par

The $n$-dimensional generalization of the Jordan system proposed in  \cite{KO21} and other $n$-dimensional reductions of the $(n+1)$-~dimensional homogeneous Euler equation considered there can be appropriate regularizing systems. The comparison of such regularizing systems and the Navier-Stokes equation for the same initial data may indicate the regimes for which the homogeneous case approximate the full case. \par

{Applications of the results obtained in the present paper to the concrete problems in physics and comparison with some previously known results 
(in particular, the numerical ones see \cite{MFNBR20}) will be discussed elsewhere. 
}
\subsubsection*{Acknowledgement} The authors are thankful to E. A. Kuznetsov for fruitful discussions. 
{We thank also the anonymous referees for useful remarks.}
This project thanks the support of the European Union's Horizon 2020 research and innovation programme under the Marie Sk{\l}odowska-Curie grant no 778010 
{\em IPaDEGAN}. We also gratefully acknowledge the auspices of the GNFM Section of INdAM under which part of this work was carried out. 
\appendix
\section{On the method of characteristics} \label{char-app}
{
Here, for convenience, we briefly recall a widely used approach for the calculation of the gradient catastrophe times, which is based on the equation (\ref{freepart-way})  (see e.g. 
\cite{L-VI,Whi,Zel70,Kuz03,BK07}),
i.e.
}
\begin{equation}
\underline{x}-\underline{x}_0 =\uu_0(\underline{x}_0) t\, .
\label{char-nBH}
\end{equation}
 The crossing condition of two characteristics passing through $\ux_0$ and $\ux_0+\underline{h}_0$
\begin{equation}
\ux-\ux_0=\uu_0(\ux_0) t\, , \qquad \ux-(\ux_0+\underline{h}_0)=\uu_0(\ux_0+\underline{h}_0) t\, .
\label{defdircat}
\end{equation}
is given by
\begin{equation}
\underline{h}_0=\uu_0(\ux_0+\underline{h}_0) t\, .
\end{equation}

For two initially close characteristics (small $|\underline{h}_0|$) the previous condition is
\begin{equation}
\ux-\ux_0-\underline{h}_0=\uu_0(\ux_0) t + \underline{h}_0 \cdot \nabla \uu_0(\ux_0) + O(|\underline{h}_0 |^2)
\end{equation}
and the intersection point satisfies 
\begin{equation}
0=\underline{h}_0 + \underline{h}_0 \cdot\nabla \uu_0(\ux_0) t+ O(|\underline{h}_0|^2)\, .
\end{equation}
In the limit of small  $|\underline{h}_0|$ the intersection point tends to the solution of
\begin{equation}
 \sum_{s=1}^n {h_0}_s \left( \delta_{is}+ \pp{}{x_s} {u_0}_i(\ux_0) t \right)= 0\, .
 \label{lefteigen}
\end{equation}
Therefore, for two infinitely close  characteristic the blow-up surface is a function of $\ux_0$ defined by the equation
\begin{equation}
t_\alpha= -\frac{1}{\lambda_\alpha(\ux_0)}\, ,
\end{equation}
where $\lambda_\alpha (\ux_0)$ are the eigenvalues of the initial data Jacobian and the vector ${h_0}_s$ defined in (\ref{defdircat}) satisfies (\ref{lefteigen}).
Consequently, the first moment of the gradient catastrophe is 
\begin{equation}
t_c= \min_{\alpha,\ux_0}\left(-\frac{1}{\lambda_\alpha(\ux_0)} \right) \equiv -\frac{1}{{\lambda_c(\ux_0}_c)}\, .
\end{equation}
The value of $\uu$ and position of the catastrophe are given by
\begin{equation}
\uu_c \equiv \uu_0({\ux_0}_c)\, , \qquad \ux_c \equiv {\ux_0}_c + \uu_c t_c\, .
\end{equation}

\end{document}